\documentclass[aps,prd,onecolumn,superscriptaddress,showpacs]{revtex4}
\usepackage{graphicx}
\usepackage{times}
\usepackage{dcolumn}
\usepackage{color}

\newcommand{\be}{\begin{equation}}
\newcommand{\ee}{\end{equation}}
\newcommand{\bea}{\begin{eqnarray}}
\newcommand{\eea}{\end{eqnarray}}

\begin{document}
%\draft

\title{Constraining Anisotropic Baryon Oscillations}

\author{Nikhil Padmanabhan}
\email{NPadmanabhan@lbl.gov}
\affiliation{Physics Division, Lawrence Berkeley National Laboratory, 1 Cyclotron Rd., Berkeley, CA 94720}

\author{Martin White}
\email{mwhite@berkeley.edu}
\affiliation{Departments of Physics and Astronomy, 601 Campbell Hall,
University of California Berkeley, CA 94720}

\date{\today}

\begin{abstract}
We present an analysis of anisotropic baryon acoustic oscillations and
elucidate how a mis-estimation of the cosmology, which leads to incorrect
values of the angular diameter distance, $d_{A}$, and Hubble parameter, $H$,
manifest themselves in changes to the monopole and quadrupole power spectrum
of biased tracers of the density field.
Previous work has focused on the monopole power spectrum, and shown that 
the isotropic ``dilation'' combination $d_{A}^{2} H^{-1}$ is robustly constrained by
an overall shift in the scale of the baryon feature. 
We extend this by demonstrating that the quadrupole power spectrum is sensitive to an
anisotropic ``warping'' mode $d_{A} H$, allowing one to break the degeneracy between
$d_{A}$ and $H$. We describe a method for measuring this warping, explicitly 
marginalizing over the form of redshift space distortions. We verify this 
method on N-body simulations and estimate that $d_A H$ can be measured
with a fractional accuracy of $\sim (3/\sqrt{V})\%$
where the survey volume is estimated in $h^{-3} {\rm Gpc}^{3}$.
\end{abstract}


\maketitle
\twocolumngrid

\section{Introduction}
Oscillations in the baryon-photon fluid prior to decoupling leave an
imprint in both the cosmic microwave background angular power spectrum
and the power spectrum of the matter with a characteristic length scale
\cite{EisReview}.
Measurements of this scale at low redshifts, for example in
large galaxy redshift surveys, provide a standard ruler for constraining
the expansion history of the Universe.
A unique feature of this method is that it can constrain both the angular
diameter distance $d_{A}(z)$ and the Hubble parameter $H(z)$ at a given
redshift, $z$.
However, the majority of current work, both theoretical and observational
\cite{SDSS_Eis,Hutsi1,Hutsi2,SDSS_Padmanabhan,
SDSS_Percival,Angulo:2007fw,Huff:2006gs,Smith:2007gi}, 
has focused on angle-averaged baryon oscillations which only
constrain the combination $d_{A}^{2}H^{-1}$.
While there have been initial efforts to use the full two-dimensional
information \cite{Okumura}, these have assumed knowledge of the
form of redshift-space distortions \cite{Kaiser,HamiltonReview}.
An open question is how to use the full two-dimensional information robustly,
i.e.~without assuming knowledge about the particular form of redshift space
distortions.

This note attempts to fill this gap. We recast the rescaling of the transverse and line-of-sight
directions as ``dilation'' and ``warping'' modes, and demonstrate that a Legendre expansion of the
two-dimensional power spectrum (or correlation function) can effectively isolate these two modes.
In particular, we show that the warping mode can be constrained by matching to a template determined
by the angle-averaged power spectrum, without any assumptions about redshift space distortions.

\section{Dilations vs Warping}

\noindent {\it Theory:} In order to understand the effects of incorrect angular and radial distance scales on the 
baryon oscillation signal, we start with the two dimensional power spectrum $P(k,\mu)$
where $k$ is the 
amplitude of the wave-vector with components $k_{\perp}$ and $k_{||}$
perpendicular and parallel to the line of sight,
and $\mu \equiv k_{||}/k$ is the cosine of the 
angle between the line of sight direction and $k$. We use the flat-sky approximation
throughout for simplicity. and 
work in Fourier space, although analogous expressions exist 
configuration space. It is also convenient to expand the angle dependence of $P(k,\mu)$
in Legendre polynomials ${\cal P}_{\ell}(\mu)$,
\begin{equation}
\label{eq:legendre}
P(k,\mu) = \sum_{\ell} P_{\ell}(k) {\cal P}_{\ell}(\mu) \,\,.
\end{equation}

An incorrect choice of $d_{A}$ and $H$ can be modeled as a deformation from a
true to observed $k_{\perp,||}$,
\begin{eqnarray}
\label{eq:}
k_{\perp} \rightarrow \alpha^{-1} (1+\epsilon) k_{\perp} \,\,\nonumber \\
k_{||} \rightarrow \alpha^{-1} (1+\epsilon)^{-2} k_{||} \,\,,
\end{eqnarray}
where $\alpha$ constrains the distance $(k_{\perp}^{2} k_{||})^{-1/3}$ or $(d_{A}^2 H^{-1})^{1/3}$, 
while $\epsilon$ constrains the distance ratio $(k_{\perp}/k_{||})^{1/3}$ or $(d_{A} H)^{-1/3}$.
This notation separates isotropic $\alpha$ deformations (``dilation'')
from anisotropic $\epsilon$ deformations (``warping'').
The measurable effect of a dilation is
to scale $k$ by $\alpha$, or slide the $P_{\ell}$ along the $k-$axis. 
Since baryon oscillations
imprint a feature with a known scale on the $\ell=0$ multipole, it is straightforward to determine $\alpha$
once this feature has been detected. Indeed, this is how baryon oscillations have been used most commonly to
date as a standard ruler. 
Furthermore, since one is matching a feature in the power spectrum (or 
correlation function), this probe is robust to a large number of potential systematics. However, it should be 
emphasized that $\alpha$ does not constrain $d_A$ and $H$ independently, but rather the combination $d_A^{2} H^{-1}$.
A dilation also changes the normalization of the power spectrum, but this is perfectly degenerate
with the bias of the mass tracer.

\begin{figure}
\begin{center}
\leavevmode
\includegraphics[width=3.0in]{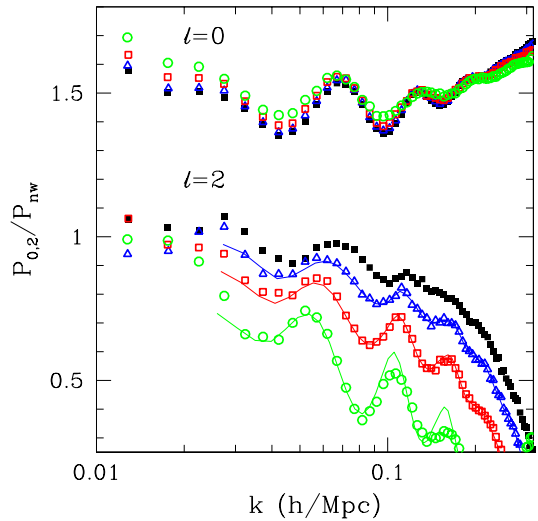}
\end{center}
\caption{The effect of warping on the $\ell=0$ and $2$ Legendre
moments of the power spectrum, measured by averaging over the 40 PM 
simulations discussed in the text. To better highlight the baryon oscillations,
we have divided out by the ``no-wiggle'' power spectrum of \cite{Nowiggle}. The 
filled squares [black] are the unwarped power spectrum, the triangles [blue] are
warped by 2\%, the squares [red] by 5\%, and circles [green] by 10\%.
The solid lines are the predictions for $\ell=2$ of the simple model in Eq.~\ref{eq:ltransform}.
The predictions for $\ell=0$ are equally accurate, although we don't plot
them to avoid cluttering the plot.
}
\label{fig:pkwarp}
\end{figure}

In order to understand the effect of $\epsilon$, we consider a toy model where 
$P_{\ell>2}=0$, a reasonable approximation on large scales.  For small
deformations $\epsilon\ll 1$ we obtain, to first
order in $\epsilon$,
\bea
k \rightarrow k \left[1 - 2\epsilon {\cal P}_{2}(\mu)\right] \,\,, \nonumber \\
\mu^{2} \rightarrow \mu^{2} - 6\epsilon \left(\mu^{2} - \mu^{4}\right) \,\,.
\eea
Substituting into Eq.~\ref{eq:legendre}, we obtain 
\bea
P_{0} \rightarrow P_{0} - \frac{2\epsilon}{5} \frac{d\,P_{2}}{d\ln k} 
    - \frac{6\epsilon}{5}P_{2} \,\,, \nonumber \\
P_{2} \rightarrow \left(1 - \frac{6\epsilon}{7}\right) P_{2}
    -\frac{4\epsilon}{7}  \frac{d\,P_{2}}{d\ln k} - 2\epsilon  \frac{d\,P_{0}}{d\ln k} \,\,,
\label{eq:ltransform}
\eea
where we ignore any hexadecapole terms. 
In the region of the baryon oscillations, the two corrections to the $\ell=0$
power have about equal magnitudes but opposite sign, resulting in an almost
exact cancellation, and therefore suppressing the correction term.
As a result, the monopole power is effectively unchanged, as one might have
naively expected; Fig.~\ref{fig:pkwarp} demonstrates this explicitly for a
suite of simulations described below.

There are three effects of warping on the $\ell=2$ component.
The first is a scaling of its amplitude.  However, as a practical matter,
the change in the amplitude will be degenerate with uncertainties in our
understanding of redshift-space distortions and therefore must be marginalized
over.  The $dP_{2}/d\ln k$ term can be recast as a scale shift in the power
spectrum, similar to the more familiar scale shift from the dilation
described above.  In principle, the existence of a feature in the power
spectrum allows one to constrain such a shift, even marginalizing over
uncertainties in the broad shape of the galaxy power spectrum.
However, Eq.~\ref{eq:ltransform} shows that the size of the shift is
suppressed by a factor of $\sim 2$.
This, combined with the fact that the amplitude of the feature is suppressed
in the $\ell=2$ component and
the larger errors for $\ell=2$ reduces the strength of statistical power of
this term (we revisit this more quantitatively below).
The final term mixes in the derivative of the $\ell=0$ component.
Normally, such a mixing would have smoothly distorted the
shape of the $\ell=2$ component, which the uncertainties in redshift space
distortions would have then forced us to marginalize over.
However, the oscillations imprint a distinctive feature in the derivative
term that one can robustly search for.
Note that the above three effects are nothing but parts of the
Alcock-Paczynski \cite{AlcockPaczynski} effect;
the only difference with its more classical application being that the
presence of a feature breaks the traditional degeneracies with broad-band
shape and amplitude.

\begin{figure}
\begin{center}
\leavevmode
\includegraphics[width=3.0in]{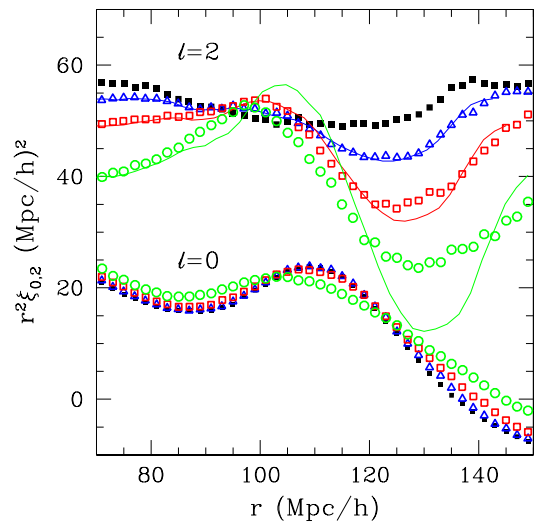}
\end{center}
\caption{As in Fig.~\ref{fig:pkwarp}, but for the redshift-space
correlation function. Note that we assume 
$\xi(r,\mu) = \xi_{0} {\cal P}_{0} - \xi_{2} {\cal P}_{2} + ...$,
so that the multipoles are mostly positive.  Note that the feature
in $\xi_2$ appears at larger separations than the comparable feature
in $\xi_0$.
}
\label{fig:xiwarp}
\end{figure}

In order to demonstrate that our toy model captures the essential
phenomenology, we consider a suite of 40 particle-mesh N-body simulations
each of $1024^3$ particles with a $2048^3$ force mesh, run assuming a
$\Lambda$CDM cosmology with $\Omega_M=0.25$, $\Omega_{\Lambda}=0.75$,
$\Omega_{b}h^{2}=0.0224$, $h=0.7$, $n=0.97$ and $\sigma_{8}=0.8$ in a 
periodic $2\,h^{-1}$Gpc box;
the precise details of these simulations will be presented elsewhere
\cite{Seoinprep}.
For definiteness we focus on the output at $z=0.3$, although this
choice was arbitrary.  We begin by considering the dark matter clustering.
For each simulation, we warp the dimensions of the volume as described
above, before computing the $\ell=0, 2$ power spectra (and correlation
functions).  The averages over the 40 boxes are shown in Figs.~\ref{fig:pkwarp}
and \ref{fig:xiwarp}.
In order to estimate the terms in Eq.~\ref{eq:ltransform} we Savitzky-Golay
filter \cite{NumRec} the unwarped power spectra with a fourth order,
nine-point filter to estimate both the smoothed power spectra and its
derivatives.  The predictions for $\ell=2$ are also plotted in
Fig.~\ref{fig:xiwarp}.
The model clearly captures the behavior of the $\ell=2$ power spectrum
when warped and is a good fit for small warping although, not surprisingly,
it starts to deviate as the warping increases.

\begin{table}
\begin{tabular}{ccc|rrr|rrr}
\hline
& & & \multicolumn{3}{c}{Dark Matter} & \multicolumn{3}{c}{Galaxies} \\
\hline
Warp & $n$ & $k_{\rm max}$ & $\langle \epsilon \rangle$ & $\sigma_{\epsilon}$  & $\chi^{2}_{\rm dof}$ 
                           & $\langle \epsilon \rangle$ & $\sigma_{\epsilon}$  & $\chi^{2}_{\rm dof}$\\
\hline
 0 & 3 &  0.2 & $ 0.0$ &  1.0 & 0.93 & $ 0.1$ &  0.9 & 0.91\\
 1 & 3 &  0.2 & $ 0.9$ &  1.0 & 0.94 & $ 1.1$ &  1.0 & 0.92\\
 2 & 3 &  0.2 & $ 2.1$ &  1.0 & 0.97 & $ 2.3$ &  1.0 & 1.01\\
 3 & 3 &  0.2 & $ 3.2$ &  1.1 & 0.94 & $ 3.2$ &  1.0 & 1.00\\
 5 & 3 &  0.2 & $ 5.4$ &  1.3 & 0.96 & $ 5.2$ &  1.0 & 1.08\\
10 & 3 &  0.2 & $ 9.9$ &  1.6 & 1.29 & $ 9.3$ &  1.1 & 1.82\\
\hline
 1 & 3 &  0.1 & $ 0.9$ &  2.1 & 1.03 & $ 1.2$ &  2.2 & 1.13\\
 1 & 3 &  0.2 & $ 0.9$ &  1.0 & 0.94 & $ 1.1$ &  1.0 & 0.92\\
 1 & 3 &  0.3 & $ 0.9$ &  1.0 & 0.69 & $ 1.8$ &  0.7 & 0.69\\
\hline
 1 & 0 &  0.2 & $ 0.9$ &  1.0 & 0.89 & $-3.9$ &  0.6 & 1.99\\
 1 & 1 &  0.2 & $ 1.0$ &  1.0 & 0.92 & $ 0.5$ &  2.0 & 0.94\\
 1 & 2 &  0.2 & $ 0.9$ &  1.0 & 0.94 & $ 1.1$ &  1.0 & 0.92\\
 1 & 3 &  0.2 & $ 0.9$ &  1.1 & 1.02 & $ 1.1$ &  1.0 & 0.98\\
\hline
\end{tabular}
\caption{\label{tab:err1} Constraining the warping $\epsilon$ of the simulation boxes using
the method outlined in the text for the dark matter and galaxies. All the warps are reported 
in per cent.
The first group of numbers demonstrate that we correctly
recover the warping, except for large warps where the model is known to be no longer accurate; 
a simple iterative procedure would work in such cases. The remaining numbers demonstrate that 
these results are insensitive to the particular choices of the bias and $k$ cut-off, except
for extremely restrictive cases.}
\end{table}

\noindent {\it Implementation:} The above discussion suggests a procedure to constrain
the warping mode. Since the $\ell=0$ mode is mostly unaffected by warping, we assume that we 
use it to correct for the isotropic shifts. Motivated by Eqs.~\ref{eq:ltransform}, we 
then fit $P_{2}(k)$ with a function of the form,
\begin{equation}
\label{eq:fitd2}
B_{n}(k) \widehat{P}_{2}\left(k\left[1-\frac{4\epsilon}{7}\right]\right) - 2\epsilon \frac{d \widehat{P}_{0}}{d\ln k} \,\,.
\end{equation}
where the hats denote undistorted quantities.
We marginalize over smooth shape distortions with a general bias function,
\begin{equation}
B_{n}(k) = \frac{\sum_{i=0}^{n} a_{i} k^{i}}{1+ \sum_{i=1}^{n} b_{i} k^{i}}.
\end{equation}
Since the $\ell=0$ power spectrum is mostly unaffected by warping, we determine
$\widehat{P}_{0}$ directly from the actual warped measurements even though Eq.~\ref{eq:fitd2} 
refers to the unwarped spectrum; the differences between the two are suppressed at 
linear order in $\epsilon$. For $\ell=2$, we suggest that $\widehat{P}_{2}$ be determined
from simulations, and allow $B_{n}(k)$ to correct for any template mismatches. In what 
follows, we determine $\widehat{P}_{2}$ and the derivative of $\widehat{P}_{0}$ by Savitzky-Golay 
filtering the unwarped average $\ell=2$ and warped $\ell=0$ power spectra respectively; the
parameters of the filter are the same as used previously.

Table~\ref{tab:err1} shows the results of fitting Eq.~\ref{eq:fitd2} to the dark matter
power spectrum in the simulations described earlier. The algorithm correctly recovers
the input warp, except as expected for the case of large warps, where Eqs.~\ref{eq:ltransform}
and ~\ref{eq:fitd2} are no longer valid. For such cases, one would simply iterate the above 
procedure.
We also note that for such large warps, the $\ell=0$ spectrum is also significantly
distorted and is therefore easily diagnosed. Table~\ref{tab:err1} also demonstrates that 
the recovered parameters are insensitive to either the form of $B_{n}$ or the cut-off in 
$k$, except when we exclude
significant amounts of the baryon oscillation signal ($k_{\rm max}=0.1 h {\rm Mpc}^{-1}$).
This suggests that our constraints are from the features in the power spectrum, and 
not the broad-band shape.

We separate the constraints from the shift in $P_{2}$ from the mixing 
of $dP_{0}/d \ln k$ by fitting for each separately in Eq.~\ref{eq:fitd2} instead of fixing 
the relationship between them. The shift is poorly constrained with an error of $\sim 8\%$,
while the error on the amplitude of the mixing term degrades to $\sim 1.2$ - $1.3$\% depending
on the particular case we consider. The two are significantly anti-correlated, 
as might be expected from Eq.~\ref{eq:ltransform}, explaining the non-trivial improvement
when they are combined.

In order to test our algorithm on a galaxy-like sample, we populate the simulations
with mock galaxies
using a local-density dependent prescription designed to match the number density
and clustering of a luminous red galaxy sample \cite{Eisenstein:2001cq}.
Generalizing the above algorithm for galaxies is trivial, but for one subtlety - the
choice of the template power spectra.
While $\widehat{P}_{0}$ can be determined by the $\ell=0$ warped galaxy spectrum, 
the form of $\widehat{P}_{2}$ must be determined from theory.
We start with the $\ell=2$ dark matter power spectrum as above, but 
suppress it with a Gaussian $e^{-(kR)^{2}}$ to account for the effect of virial motions within halos.
We do not simultaneously fit for the warp and  
$R$, but fix $R$ to $5 h^{-1} {\rm Mpc}$ appropriate to our particular sample of galaxies,
and allow $B_{n}(k)$ to correct for errors in the suppression term. 
In practice, one would determine the form of this suppression 
by fitting to the small-scale galaxy correlation function. The results are summarized in 
Table~\ref{tab:err1}; as with the dark matter, we see that our algorithm correctly recovers the
input warps, except in two cases. The method fails for $k_{\rm max} = 0.3 h {\rm Mpc}^{-1}$, since 
(for this particular galaxy sample) smaller scales are significantly affected by virial
motions and Eq.~\ref{eq:fitd2} is no longer a good description of the 
power spectrum. In principle, this could be alleviated by e.g.\ collapsing clusters to
remove the virial motions, or down-weighting these scales. 
Similarly, the method fails if one chooses too restrictive a form for
$B_{n}(k)$;
this is seen most spectacularly for $n=0$, which assumes no scale dependence to the bias,
although the best fit is disfavored in this case at greater than $99.9\%$.
Increasing $n$ quickly corrects this problem, and we correctly recover the input warps.
Both these cases emphasize the need to choose a flexible form to describe the 
connection between galaxies and dark matter. We have demonstrated that a simple choice
works in the case considered here, but more comprehensive investigations
must be the subjects of future work.

\noindent {\it Comparison to Fisher matrix estimates:}
A Fisher matrix calculation \cite{SeoEis07} predicts that measuring 
baryon oscillations in a $8\,h^{-3}{\rm Gpc}^{3}$ volume yields 
fractional errors
on $d_{A}$ and $H$ of 0.7\% and 1.8\% respectively with a cross-correlation
coefficient of $0.4$. This implies errors on $\alpha$ and $\epsilon$ of $0.6\%$
and $0.7\%$ with a cross-correlation coefficient of $0.4$.
The Fisher error on $\epsilon$ is $\sim 40\%$ better than our estimates.
This is not too surprising, since the Fisher matrix calculation of
\cite{SeoEis07} makes optimal use of the entire power spectrum assuming a
perfect parameterized model (and also requires a subtraction of information
{}from broad-band power). The degradation of dark energy constraints will be
smaller than this, since the dominant constraint comes from the
dilation mode; the precise degradation will depend on the 
redshift range of the experiment and the particular priors used.

\section{Summary}
The standard ruler method for constraining the expansion history of the
Universe, as exemplified by baryon acoustic oscillations, can measure
both $d_A(z)$ and $H(z)$ with high accuracy.
However, the most commonly used angle-averaged measures constrain
only the combination $d_A^2H^{-1}$.  We have shown that by measuring both
the monopole and quadrupole power spectra $P_{0,2}$, it is possible to break this
degeneracy and measure $d_A$ and $H$ independently without assuming
a form for redshift space distortions.  

By modeling the manner in
which ``warping'' modes modify $P_{2}$, we have given an explicit method 
for measuring $d_AH$, allowing us to break the degeneracy between $d_{A}$ and
$H$ from the angle-averaged measurement. We have 
demonstrated this method on N-body simulations and estimate
that the combination $d_{A} H$ can be constrained with an accuracy of
$\sim (3/\sqrt{V})\%$
where the survey volume is estimated in $h^{-3} {\rm Gpc}^{3}$.
Since our method relies only on the information contained in the acoustic
feature, it is insensitive to the details of the complex relation between
mass and galaxies or the specific form of the redshift space distortions.
This degrades the constraint on $d_{A} H$ by $\sim 40\%$ compared with traditional 
Fisher matrix estimates which assume perfect knowledge of redshift space distortions 
(even when they marginalize over the shape of the power spectrum). 
This is, of course, a conservative assessment, and improvements in our
abilities to robustly the relationship between galaxies and matter will 
improve constraints.

\acknowledgments
We thank Daniel Eisenstein and Will Percival for discussions that
helped shape our thinking on this problem, and Eric Linder
for comments on an earlier draft. NP also thanks NYU and
Harvard for visits during which significant portions of this work were completed.
NP is supported by NASA Hubble Fellowship 
NASA HST-HF-01200.01 and an LBNL Chamberlain Fellowship.
MW is supported by NASA. This project used the computing
resources at the National Energy Research Scientific Computing Center.
This work was supported by the Director, Office of Science, of the U.S. 
Department of Energy under Contract No. DE-AC02-05CH11231.

%\bibliography{biblio,preprints}

\bibliography{anibao}

\end{document}